\documentclass[aps,pra,groupedaddress,nofootinbib,notitlepage,showpacs,floatfix]{revtex4}
\usepackage{amsfonts}
\usepackage{amsmath}
\usepackage{graphicx}
\usepackage[normalem]{ulem}
\usepackage{times}
\usepackage{amssymb}
\usepackage{hyperref}
\usepackage{textcomp}
\usepackage{color}
\usepackage{xcolor}
\usepackage{pdfpages}
\usepackage{pstricks,epstopdf}
\usepackage[utf8]{inputenc}

\begin{document}
\title{Enhanced violation of a Leggett-Garg inequality under nonequilibrium thermal conditions}
\author {Juan C. Castillo}
\affiliation{Departamento de F\'{i}sica, Universidad de los Andes,
A.A. 4976 Bogot\'a, D.C.,Colombia}
\author {Ferney J. Rodr\'{i}guez }
\affiliation{Departamento de F\'{i}sica, Universidad de los Andes,
A.A. 4976 Bogot\'a, D.C.,Colombia}
\author {Luis Quiroga}
\affiliation{Departamento de F\'{i}sica, Universidad de los Andes,
A.A. 4976 Bogot\'a, D.C.,Colombia}
\date{\today}

\begin{abstract}
We investigate both analytically and numerically violations of a Leggett-Garg inequality (LGI) for a composite quantum system in contact with two separate reservoirs at different temperatures. Remarkably we find that LGI violations can be enhanced when a heat current is established at low temperatures in a steady-state regime. Based on a Kraus operator decomposition of the non-unitary evolution for a system formed by two interacting spins or quantum bits, we provide analytical support for power law relations between dissipation strength and mean temperature in the borderlines separating parameter regions where non-equilibrium conditions affect differently the maximal LGI violation. Furthermore, a correspondence between spatial and temporal correlation inequalities is shown to persist even in such nonequilibrium thermal settings.

\end{abstract}

\maketitle

\section{Introduction}
Since the seminal work by John Bell ~\cite{bell} about quantum spatial correlations, an ever increasing number of experiments has been devoted to test the foundations of quantum mechanics. In a parallel way, and triggered by those tests, practical schemes for conveying quantum information have flourished. New aspects of the fascinating world of quantum correlations arose when temporal, instead of spatial, correlations were proposed by Leggett and Garg \cite{leggett}, and subsequently by many other authors \cite{paz,huelga,kofler}. In those works
different inequalities between two-time correlations have been proposed that should hold whenever a classical description is valid. An emerging Leggett-Garg inequality (LGI) violation would mark a borderline between the classical and quantum worlds. Of special interest has been
the guide provided by LGI for testing macroscopic realism in condensed matter systems \cite{nori,palacios,goggin,knee,jordan,athalye,waldherr} as well as in quantum optics setups \cite{xu,dressel}
where focus has been put on LGI violations for open quantum systems in contact with realistic reservoirs.

On the other hand, one of the most important challenges to
detect superposition of quantum macroscopic states is the fragility of these states caused by decoherence effects. Despite the ubiquitous presence of non-equilibrium situations in quantum mesoscopic and macroscopic physics, few formal results about
quantum correlations in such regimes are available \cite{quiroga,polko,cazalilla,plenio}.
Many efforts encompassing a large variety of physical systems have
been directed to study the effects of spatial correlations, as measured by the concurrence \cite{brandes,quiroga,petruccione1} and quantum discord (QD) \cite{wu1}, under non-equilibrium gradients.
However, LGI studies have been very scarce in systems out of thermodynamic
equilibrium, although time correlations in condensed matter systems show complex and interesting behaviors, like those observed in ultracold atom systems \cite{greiner}, inductively
(capacitively) coupled flux (charge) superconducting qubits \cite{simmonds},
and even energy transfer processes in photosynthetic purple bacteria \cite{sarovar}.

How do thermal non-equilibrium conditions affect quantum two-time correlations?  From a theoretical point of view important insights have been given
in Ref.\cite{scully} where nanothermodynamics relevance of quantum coherences has been addressed. From the experimental side Ref.\cite{giazotto} shows an elegant experiment where thermal and electric currents interfere in a Josephson junction. Here, we propose a set up where the relationship between two-time quantum correlations and out-of-equilibrium thermal conditions can be systematically explored through the LGI violations.
Here, we examine a LGI for a two-qubit system and show that violations are enhanced when a heat current, ${\cal J}$, is flowing through it.

For nonequilibrium dynamics, what is the
optimal relationship between dissipation strength ($\Gamma$) and temperature range ($T$) which allows the detection of quantum correlations through macroscopic transport measurements?
The theoretical approach we provide, in terms of Kraus operator dynamics evolution, gives a
general framework to identify non-classical features through LGI violations for quantum systems in contact with two memoryless (Markovian) separate reservoirs at different temperatures. We find that dissipation, inter-qubit interactions and mean temperature compete, resulting in a rich $T$-$\Gamma$ phase diagram where power-law relations mark the boundaries between different zones where LGI violations are either absent, decreasing or increasing with an applied temperature gradient.

The structure of the paper is as follows. In Sec~\ref{sec:II} we
give the details about the Leggett-Garg inequalities in a generic non-thermal steady state and
the calculation of two-time quantum correlations. In Sec.~\ref{sec:III}
we describe our model of two interconnected qubits each one interacting with an independent thermal bath. The nonequilibrium situation arises when the two baths are at different temperatures.
In Sec.~\ref{sec:IV}, we present our main analytical and numerical results about the nonequilibrium enhanced LGI violation.
Our main conclusions are summarized in Sec.~\ref{sec:V} while technical details are described
in the Appendixes.

\section{Legget-Garg inequalities in a nonequilibrium steady-state}
\label{sec:II}
The two-time correlation inequality, LGI, extends the well-known Bell inequality (designed to probe spatial quantum correlations) to the time domain. One form the LGI can take is \cite{leggett,paz,huelga,avis}:
\begin{eqnarray}
F(t_1,t_2,t_3)=C(t_1,t_2)+C(t_2,t_3)-C(t_1,t_3)\leq 1 \label{Eq:5}
\end{eqnarray}
where $C(t_i,t_j)$ is the two-time correlation of a dichotomic observable $\hat{Q}$ (eigenvalues $a$=$\pm1$)
between times $t_i$ and $t_j$ and $t_1 < t_2 < t_3$. To evaluate  $F(t_1,t_2,t_3)$ in Eq.(\ref{Eq:5}),
a Kraus operator $\hat{K} (t)$ approach has been used \cite{krauss}. For the sake of completeness, here we briefly review the Kraus operator formalism and show how it applies to a nonequilibrium steady-state regime.

\subsection{Kraus operators}
Consider a physical system with quantum states in a Hilbert space $\mathcal{H}$. The time evolution of its density matrix $\rho(t)$ is described by the following Lindblad equation:
\begin{equation}\label{eq:ecDif}
 \frac{\partial}{\partial t}\rho(t)=\mathcal{L}\left[\rho(t)\right]=-i\left[H,\rho(t) \right]+\mathcal{D}\left[\rho(t)\right]
\end{equation}
with a time-independent Lindbladian superoperator $\mathcal{L}$, where $\mathcal{D}$ is the dissipator or nonunitary evolution term. The goal is to find a form for the evolution superoperator $\mathcal{E}(t_1-t_2)$, which maps an initial density operator of the system to a final density operator, as defined by

\begin{equation}\label{eq:defEvolucion}
 \rho(t_2)=\mathcal{E}(t_1-t_2)\left[\rho(t_1)\right]
\end{equation}
This form generates a semigroup associated with a Markovian dynamics. Thus, the evolution only depends on the time interval between $t_1$ and $t_2$. An elegant and efficient form of expressing $\mathcal{E}(t_1-t_2)$ is using the Kraus operators via the following equation:
\begin{equation}
\rho(t'+t)=\mathcal{E}(t)\left[\rho(t')\right]=\sum_{\mu}K_\mu(t)\rho(t')K_\mu^\dagger(t)
\end{equation}
where the Kraus operators $K_\mu$ fulfill the condition $\sum_\mu K_\mu^\dagger(t) K_\mu(t)=1$ in order to preserve the unit trace of the density operator at any time. It can easily be shown that, replacing the Kraus operators with unitary evolution operators, the exact results for closed quantum systems are retrieved. Since the Kraus representation describes equivalently the nonunitary time evolution of an open quantum system we can conclude that the Kraus operators depend only on the time interval and give direct access to time correlation functions of the open system. However, the explicit calculation of Kraus operators can be a complex task beyond simple, and generally, non-interacting composite quantum systems. As discussed below we find them in a numerical way for an interacting spin chain out-of-equilibrium.

\subsection{Two-time quantum correlations in terms of Kraus operators}
The time correlations $C(t_i,t_j)$ in Eq.(\ref{Eq:5}) are evaluated as \cite{leggett,kofler}
\begin{eqnarray}
\label{Eq:6}
C(t_i,t_j)&=&p(^+t_i)q(^+t_j|^+t_i)+p(^-t_i)q(^-t_j|^-t_i)
 \nonumber
 \\
 &-&p(^-t_i)q(^+t_j|^-t_i)-p(^+t_i)q(^-t_j|^+t_i) \label{Eq:6}
\end{eqnarray}
where $p(^at_i)$ is the probability of obtaining the result $a=\pm 1$ at $t_i$, and $q(^at_i|^bt_l)$ is the conditional probability of getting the result $a=\pm 1$ at $t_i$ given that result $b=\pm 1$ was obtained at $t_l$.

Now, we show the connection between Kraus operators and conditional/unconditional probabilities involved in obtaining two-time quantum correlations in Eq.(\ref{Eq:6}).
To find each one of these probabilities, we must use the projector $\hat{\Pi}^{\pm}$, related to the eigenspaces of the observable $\hat{Q}$, in terms of which $p(^at_i)=\mathrm{Tr}\left\{ \hat{\Pi}^a\hat{\rho}(t_i)\right\}$. To proceed further, the conditional probabilities $q(^at_i|^bt_l)$ should be obtained. A generic term in the right hand side of Eq.(\ref{Eq:6}) becomes
\begin{eqnarray}
p(^at_i) q(^at_i|^bt_l)&=&\mathrm{Tr}
\left\{
\hat{\Pi}^a\sum_{\nu,\mu} \hat{K}_\nu (t_i-t_l)\hat{\Pi}^b \hat{K}_\mu(t_l)\hat{\rho}_0
\right.
  \nonumber
  \\
& & \left .
\hat{K}^\dagger_\mu(t_l) \hat{\Pi}^b \hat{K}^\dagger_\nu (t_i-t_l)\right\} \label{Eq:98}
\end{eqnarray}

The expression for $p(^jt_i)$ (with $j=\pm$) is now

\begin{eqnarray}
 p(^jt_i)=\mathrm{Tr}\left\{ \Pi^j\rho(t_i)\right\}=\mathrm{Tr}\left\{ \Pi^j\sum_\mu K_\mu (t_i) \rho_0K^\dagger_\mu (t_i)\right\} \label{Eq:probMixto}
\end{eqnarray}

To obtain an expression for $q(^jt_i|^kt_l)$ (where $j,k=\pm$), we must also find the density matrix corresponding to a measurement result $k$ obtained at time $t_l$:

\begin{equation}
 \rho^k(t_l)=\frac{\Pi^k\rho(t_l)\Pi^k}{\mathrm{Tr}\left\{\Pi^k\rho(t_l)\right\}}=\frac{\Pi^k\sum_\mu K_\mu (t_l) \rho_0K^\dagger_\mu (t_l)\Pi^k}{p(^jt_l)}
\end{equation}

This state now evolves until $t_i$, when the state of the system is $\sum_\nu K_\nu (t_i-t_l)\rho^k(t_l)K^\dagger_\nu (t_i-t_l)$, so that

\begin{equation}
 q(^jt_i|^kt_l)=\frac{\mathrm{Tr}\left\{\Pi^j\sum_{\nu\mu} K_\nu (t_i-t_l)\Pi^k K_\mu(t_l)\rho_0 K^\dagger_\mu(t_l) \Pi^k K^\dagger_\nu (t_i-t_l)\right\}}{p(^jt_i)}
\end{equation}

With all these ingredients the two-time quantum correlations can finally be written as (see Appendix \ref{app:Atimedepquancorr-details} for details):

\begin{eqnarray}
 C(t_1,t_2)=1-2p(^+t_1)-2p(^+t_2)+4\mathrm{Re}\left[g(t_1,t_2)\right] \label{Eq:correlacionMixto}
\end{eqnarray}
This expression depends mainly of the correlation function,
\begin{eqnarray}
g(t_l,t_i)=\mathrm{Tr}\left\{\Pi^+\sum_{\nu} K_\nu (t_i-t_l) \Pi^+ \rho(t_l) K^\dagger_\nu (t_i-t_l)  \right\} \label{Eq:gMixto}
\end{eqnarray}

Based on the correlations $C(t_i,t_j)$, the LGI function $F(t_1,t_2,t_3)$ can be written as
\begin{eqnarray}
F(t_1,t_2,t_3)=C(t_1,t_2)+C(t_2,t_3)-C(t_1,t_3)\leq 1\\
F(t_1,t_2,t_3)=1-4p(^+t_2)+4Re\left[g(t_1,t_2)+g(t_2,t_3)-g(t_1,t_3)\right]\leq 1
\label{Eq:LG1General}
\end{eqnarray}

This last equation, jointly with Eq.(\ref{Eq:gMixto}), is our main result allowing us to explore LGI violations in nonequilibrium set ups.
From now on, we limit ourselves to consider a long-time limit or a steady-state situation where the density operator will be denoted as $\hat{\rho}_{ss}=\hat{\rho}(t\rightarrow \infty)$ which fulfills
\begin{equation}
\dot{\rho}_{ss}=\mathcal{L}\rho_{ss}=0
\end{equation}
or equivalently
\begin{equation}
\mathcal{E}\left[t\right](\rho_{ss})=\sum_\mu K_\mu (t) \rho_{ss}K^\dagger_\mu (t)=\rho_{ss}\quad \forall t
\end{equation}
Due to this stationary state assumption, 
two-time correlations depend only on the time difference between two instants, which implies
the non-invasive measurement requirement in LGI settings \cite{huelga}.

In the calculation of the LGI function, $F(t_1,t_2,t_3)$, all three times $t_1$, $t_2$ and $t_3$ are taken in the steady-state regime and we fix them as $t_1=0$, $t_2= \tau$ and $t_3=2 \tau$. Let us now find a simplified expression for the LGI function when the system is in the steady-state. The first step is to find an expression for $p(^jt_i)$ by using Eq.(\ref{Eq:probMixto}):

\begin{equation}
p(^jt_i)=\mathrm{Tr}\left\{ \Pi^j\sum_\mu K_\mu (t_i) \rho_{ss}K^\dagger_\mu (t_i)\right\}=\mathrm{Tr}\left\{ \Pi^j\rho_{ss}\right\}
\end{equation}
We can thus see that it is time independent, and therefore we will simply call it $p^j$. We must also find an expression for $g(t_l,t_i)$ from Eq.(\ref{Eq:gMixto}):

\begin{align}\label{eq:gEstadoEstacionario}
g(t_l,t_i)=\mathrm{Tr}\left\{\sum_{\nu\mu} \Pi^+ K_\nu (t_i-t_l) \Pi^+ K_\mu(t_l)\rho_0 K^\dagger_\mu(t_l)  K^\dagger_\nu (t_i-t_l)  \right\} \nonumber \\
=\mathrm{Tr}\left\{\sum_{\nu\mu} \Pi^+ K_\nu (t_i-t_l) \Pi^+\rho_{ss}  K^\dagger_\nu (t_i-t_l)  \right\}
\end{align}
The last expression implies that $g(t_l,t_i)$ depends only on the difference between $t_i$ and $t_l$, so we will call it $g(t_i-t_l)$. The expression for correlations, based on Eq.(\ref{Eq:correlacionMixto}), is now

\begin{equation}\label{eq:corrEstadoEstacionario}
 C(t_1,t_2)=1-4p^++4\mathrm{Re}\left[g(t_2-t_1)\right]
\end{equation}
which also depends only on the time difference. This result is as expected, since if the system is already in the steady state at $t_1$, it is only relevant the time interval between the two times at which the correlation is evaluated.
Finally, the LGI function in Eq.(\ref{Eq:5}) can be written as
\begin{eqnarray}
F(0,\tau,2\tau)=1-4p^+ +4\mathrm{Re}\left[ 2g(\tau) - g(2\tau)\right])\leq 1
\label{Eq:8}
\end{eqnarray}
where $p^+=Tr\left\{\hat{\Pi}^+\hat{\rho}_{ss}\right\}$ denotes the long-time limit of the probability for obtaining the result $a=+1$, $\hat{\Pi}^+$ the corresponding projector and
\begin{eqnarray}
g(\tau)=
\mathrm{Tr}\left\{\hat{\Pi}^+\sum_{\nu} \hat{K}_\nu (\tau) \hat{\Pi}^+ \hat{\rho}_{ss} \hat{K}^\dagger_\nu (\tau)  \right\} \label{eq:gMixto}
\end{eqnarray}
This last expression is of the highest importance for the present work as it is the responsible for producing LGI violations.

\section{The model}
\label{sec:III}
\subsection{The Hamiltonian}
We consider a bipartite quantum system in contact with separate thermal baths at different temperatures $T_1$ and $T_2$. To be specific, the quantum
system of interest is a two spin chain described by:
\begin{equation}
\hat{H}_Q=\frac{\epsilon_1}{2}\hat{\sigma}_{z,1}+\frac{\epsilon_2}{2}{\sigma}_{z,2}+ V(\hat{\sigma}_{1}^{+}\hat{\sigma}_{2}^{-}+
\hat{\sigma}_{1}^{-}\hat{\sigma}_{2}^{+}) \label{Eq:2}
\end{equation}
where $\epsilon_i$ is the energy splitting of the i-th qubit, $\hat{\sigma}^{\pm}_{z,i}$ denote Pauli matrices and $V$ describes the inter-qubit interaction strength. The baths
are represented by sets of harmonic oscillators with Hamiltonians $\hat R_i=\sum_n\Omega_{n,i} \hat{a}^{\dag}_{n,i}\hat{a}_{n,i}$ where $\hat{a}^{\dag}_{n,i}$ ($\hat{a}_{n,i}$) creates (destroys) an excitation in the i-th reservoir while the coupling of each qubit with its separate reservoir is given by
$\hat{H}_{int,i}=\hat{\sigma}_{i}^{+}\sum_n g_n^{(i)}\hat{a}_{n,i}+\hat{\sigma}_{i}^{-}\sum_n g_n^{(i)}\hat{a}_{n,i}^{\dagger}$ where $g_n^{(i)}$ denote the system-bath coupling strengths. In the Born-Markov framework the system-bath couplings are essentially determined by the bath spectral density of the form
 $J_i( \omega)=\Gamma n_i(\omega)$ where $\Gamma$ denotes the coupling strength (taken identical for both reservoirs) and $n_i(\omega)=(e^{\beta_i\omega}-1)^{-1}$, the Bose-Einstein distribution of excitations in the $i$-th bath at inverse temperature $\beta_i$  \cite{quiroga}. For the sake of simplicity we limit ourselves to the {\it symmetric-qubit} case, i.e. $ \epsilon_1= \epsilon_2= \epsilon$. In the following, energy, frequency and temperature will be measured in units of the inter-qubit interaction strength $V$ while time will be expressed in units of $ V^{-1}$ by letting $\hbar=K_B=1$. We thus set $\tilde{\epsilon}=\epsilon/V$, $\tilde{\Gamma}=\Gamma/V$, $\tilde{\omega}=\omega/V$, $\tilde{T}=T/V$, and $\tilde{\tau}=V\tau$.

The eigenstates for the qubit Hamiltonian in Eq.(\ref{Eq:2}) are given, in the $\hat{\sigma}_{z,i}$ basis, by: $|1\rangle=|+,+\rangle,|2\rangle=(|+,-\rangle-|-,+\rangle)/\sqrt{2},
|3\rangle=(|+,-\rangle+|-,+\rangle)/\sqrt{2}$ and
$|4\rangle=|-,-\rangle$. The non-equilibrium thermal steady-state density operator for the pair of qubits turns out to be diagonal in this basis taking the form of a direct product as \cite{quiroga}
\begin{eqnarray}
\hat{\rho}_{ss}=\hat{\rho}^{ss}_1\otimes \hat{\rho}^{ss}_2=\sum_{\alpha=1}^{4} c_{\alpha}|\alpha\rangle\langle \alpha| \label{Eq:q99}
\end{eqnarray}
where $\hat{\rho}^{ss}_j$ denotes the steady-state density operator of a {\it single fictitious} qubit of energy splitting $\tilde \omega_j$ in thermal equilibrium with a single reservoir at an effective temperature corresponding to the mean number of thermal excitations as given by $n(\tilde \omega)=\frac{1}{2}\sum\limits_{i=1,2}n_i(\tilde \omega)$ with index $i$ representing each separate thermal bath. The two effective qubits have energy splittings given by $\tilde\omega_1=|\tilde\epsilon-1|$ and $\tilde\omega_2=\tilde\epsilon+ 1$. Thus, $\hat{\rho}^{ss}_j=\mathrm{diag}\{\frac{n(\tilde\omega_j)}{2n(\tilde\omega_j)+1}, \frac{n(\tilde\omega_j)+1}{2n(\tilde\omega_j)+1}\}$ implying that the steady-state spin chain density operator in Eq.(\ref{Eq:q99}) does not depend on the system-bath coupling strength $\Gamma$.
Of special interest for the discussion of results below are the coefficients $c_2$ and $c_3$ in Eq.(\ref{Eq:q99}), given by $c_2=\left(\frac{n(\tilde\omega_1)}{2n(\tilde\omega_1)+1}\right) \left(\frac{n(\tilde\omega_2)+1}{2n(\tilde\omega_2)+1}\right)$ and $c_3=\left(\frac{n(\tilde\omega_1)+1}{2n(\tilde\omega_1)+1}\right) \left(\frac{n(\tilde\omega_2)}{2n(\tilde\omega_2)+1}\right)$, which determine the heat current a cross the quantum system by the annihilation of an excitation from one reservoir and the subsequent creation of another excitation in the other reservoir through a flip-flop process in the two-spin chain. Notice that in the special case $\tilde\epsilon=1$, one of the fictitious qubits has zero splitting, $\tilde\omega_1=0$, which is equivalent to having at least one bath at infinite temperature since $n(\tilde\omega)\rightarrow \infty$.
In order to assess under which conditions the LGI can be violated, $F(0,\tilde\tau,2\tilde\tau) > 1$ in Eq.(\ref{Eq:8}), we pick the {\it single} qubit operator $\hat{Q}=\hat{\sigma}_{z,1}$ as the dichotomic observable to evaluate the LGI. The projector to the
$|+\rangle$ eigenspace for this observable is $\hat{\Pi}^+  = |+\rangle \langle + | \otimes \hat{\mathbb{I}}=|1\rangle\langle1|+\frac{1}{\sqrt{2}}\sum_{\alpha=2}^{3}|\alpha\rangle\langle \alpha|$.
\subsection{General considerations}
To gain physical insight on quantum two-time correlations and LGI violations for a nonequilibrium thermal spin chain, we first consider a simple case: once the spin chain reaches the nonequilibrium steady-state regime, we proceed to break its couplings to the heat reservoirs and let it evolve unitarily under the action of Hamiltonian $\hat{H}_Q$. In this case, the LGI-$F$ function can be analytically evaluated, yielding to
\begin{eqnarray}
F(0,\tilde\tau,2\tilde\tau)&=&1+r_{23}\left[2\cos{(2\tilde\tau)} - \cos{(4\tilde\tau)}-1 \right] \label{Eq:200}
\end{eqnarray}
which only involves, through $r_{23}=c_2+c_3$, the density matrix elements corresponding to eigenstates $|2\rangle$ and $|3\rangle$, as expected. Under the sole action of $\hat{H}_Q$ the states $|1\rangle$ and $|4\rangle$
are thus decoupled from the other two states. It is then easily seen that $F(0,\tilde\tau,2\tilde\tau)$ begins at a value of 1 for $\tilde\tau=0$. The LGI is violated during an initial time interval from $\tilde\tau=0$ to $\tilde\tau=\pi/4$, reaching a maximal violation when $\tilde\tau=\pi/6$ irrespective of the bath temperatures. However, the maximum value of the LGI violation (MLGI), as defined by $MLGI=F^{(Max)}(0,\tilde\tau,2\tilde\tau)-1\geq 0$, does depend on $r_{23}$ and consequently on the bath temperatures.

Now, we restore the couplings of the spin chain with the baths which makes it imperative to evaluate the quantum open system's Kraus operators numerically (details will be presented elsewhere). Using this procedure, the two-time correlation function $C_N(\tilde\tau)$ can be safely approximated by a fitting correlation function of the form $C_F(\tilde\tau)= 1-r_{23}\left[1-e^{-\tilde\Gamma_V \tilde\tau}\cos{(2\tilde\tau)}\right]$, where $\frac{\tilde\Gamma_V}{\tilde\Gamma}=
\sum\limits_{j=1}^2\left[2n(\tilde\omega_j)+1\right]$. The fit is good enough that $|C_N(\tilde\tau)-C_F(\tilde\tau)|\leq 10^{-5}$ for the full numerical results in the range $\tilde\tau < 6$, validating our analytical
results described below.
Thus, by using the fitting correlation functions, $C_F(\tilde\tau)$, the LGI in Eq.(\ref{Eq:8}) can be written as
\begin{eqnarray}
\label{Eq:fleget}
&  &F(0,\tilde\tau,2\tilde\tau)=1+r_{23}\\
\nonumber
& & \left[2e^{-\tilde\Gamma_V \tilde\tau} \cos{(2\tilde\tau)} -e^{-2\tilde\Gamma_V \tilde\tau}\cos{(4\tilde\tau)}-1 \right]
\leq 1
\end{eqnarray}
In contrast to the steady-state nonequilibrium thermal concurrence \cite{quiroga,petruccione1} and QD \cite{wu1}, the LGI-$F(0,\tilde\tau,2\tilde\tau)$
depends explicitly on the dissipation rate $\tilde\Gamma$. It is of special interest to relate the LGI-$F$ function with a measurable transport quantity such as the heat current flowing through the spin chain in response to a temperature gradient. An analytical expression for the heat current under a temperature gradient, $\Delta {\tilde T}={\tilde T}_1-{\tilde T}_2$, as calculated from
${\cal J}(\Delta \tilde T) = \mathrm{Tr}\left\{ \hat{H}_{Q} \hat{{\cal L}}_1\right\}$ can be found in Ref.\cite{quiroga}.
Although a closed equation relating $F(0,\tilde \tau,2\tilde \tau)$ with the heat flow is possible, the expression is cumbersome and will be skipped here, since their relationship is still better appreciated by looking at the figures  as discussed below.

\begin{figure}[tbh]
\includegraphics[height=7.0cm,width=8.2cm]{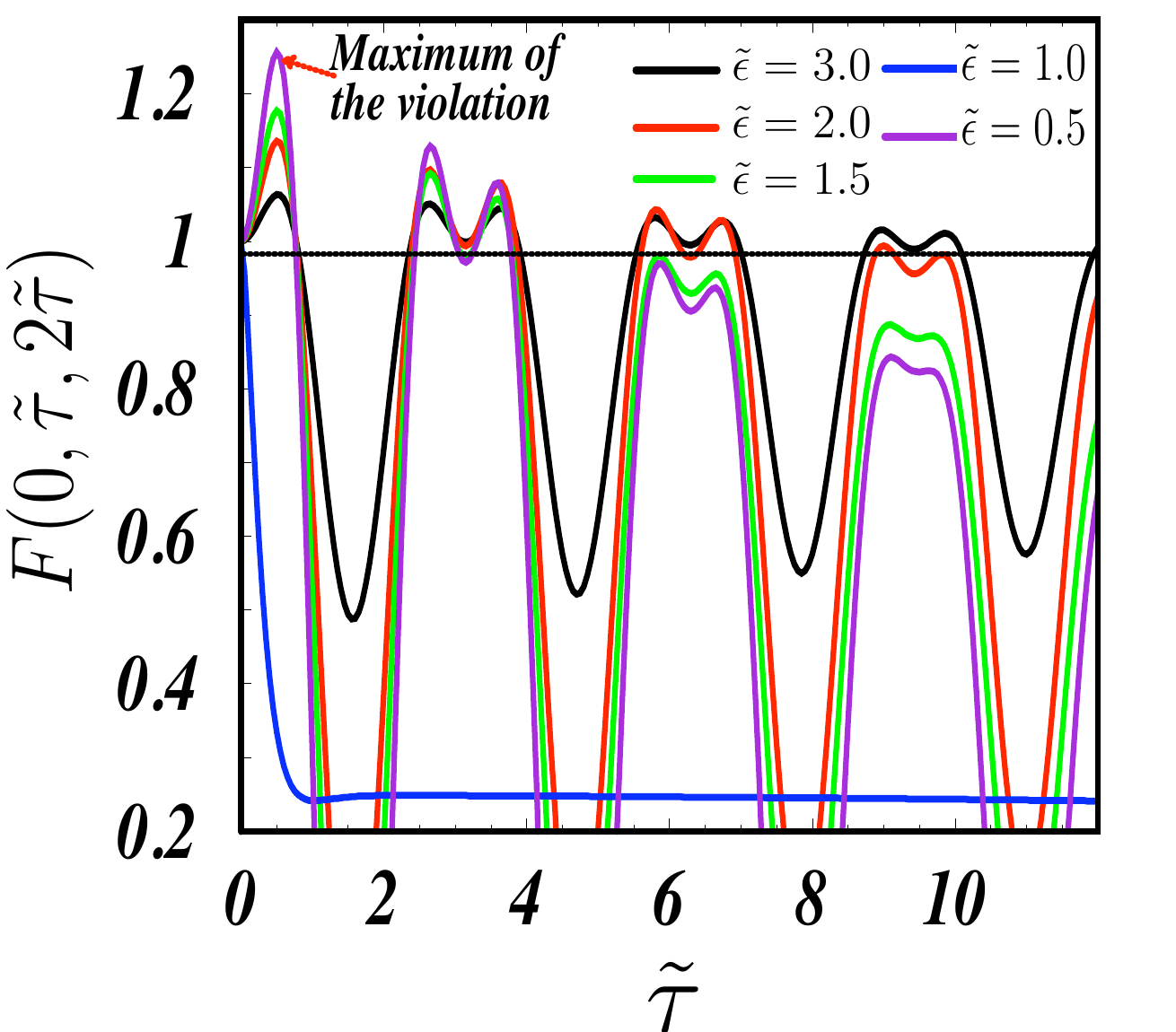}
\caption{(Color online). LGI-F function at thermal equilibrium, i.e. $\Delta \tilde T=0$. $F(0,\tilde \tau,2\tilde \tau)$ for symmetrical qubits with different splittings and identical bath temperatures $\tilde T_1=\tilde T_2=1.1$. The LGI is violated whenever $F(0,\tilde \tau,2\tilde \tau) > 1$.}
\label{fig1fvst}
\end{figure}

\section{Results and discussions}
\label{sec:IV}
Although the main idea behind the initial proposal of the LGI was to test macroscopic realism \cite{leggett}, we will use LGI with a different purpose. As we saw in the discussion leading to LGI, this inequality is never violated in the time evolution of classical systems. Therefore, any system which exhibits any degree of violation to the LGI has some fundamental quantum behavior. We can therefore use LGI violations, similarly to the use of the entanglement of formation and the quantum discord, to identify systems which evolve in some no classical way. This is the reasoning line we will follow from now on.

To illustrate the behavior of the MLGI violation under nonequilibrium thermal conditions we proceed to consider realistic parameters. The set of chosen parameters (in V units), for the results we discuss below, are in the actual experimental range for superconductor qubits $\tilde \epsilon_i\sim 5 $ and $\tilde \Gamma\sim 10^{-3} $ \cite{simmonds} while for excitons in biological photosynthetic systems $\tilde \epsilon_i\sim 1.4 $ and
$\tilde \Gamma\sim 0.4  $ \cite{sarovar}. First, we analyze results for the equilibrium case, i.e. $ \Delta \tilde T=0$. As plotted in  Fig. \ref{fig1fvst}, violations of LGI are still visible up to $\tilde \tau=10$ for an equilibrium temperature $\tilde T_M=\tilde T_1=\tilde T_2=1.1$.
For weak inter-qubit couplings ($\tilde \epsilon > 1$) the violation of the LGI is small compared with the strong inter-qubit coupling result in the range of $0 <\tilde  \tau < 1$. However, for $1 <\tilde  \tau < 5$ that behavior is
reversed and the classical non-violation regime prevails ($ F(0,\tilde \tau,2\tilde \tau) < 1$). Moreover,
in the weak inter-qubit coupling the LGI violation stands for longer times.
By decreasing the qubit splitting $\tilde \epsilon$ the LGI violation is becomes large for
short times, as can be seen
in Fig.\ref{fig1fvst}. For the special case $\tilde \epsilon=1$, the LGI violation breaks down for any $\tilde \tau$ and the classical behavior emerges, as should be for a quantum system in contact with an infinite temperature bath as pointed out above.

\begin{figure}[tbh] 
\hspace{-1.0cm}
\includegraphics[height=4.0cm,width=3.8cm]{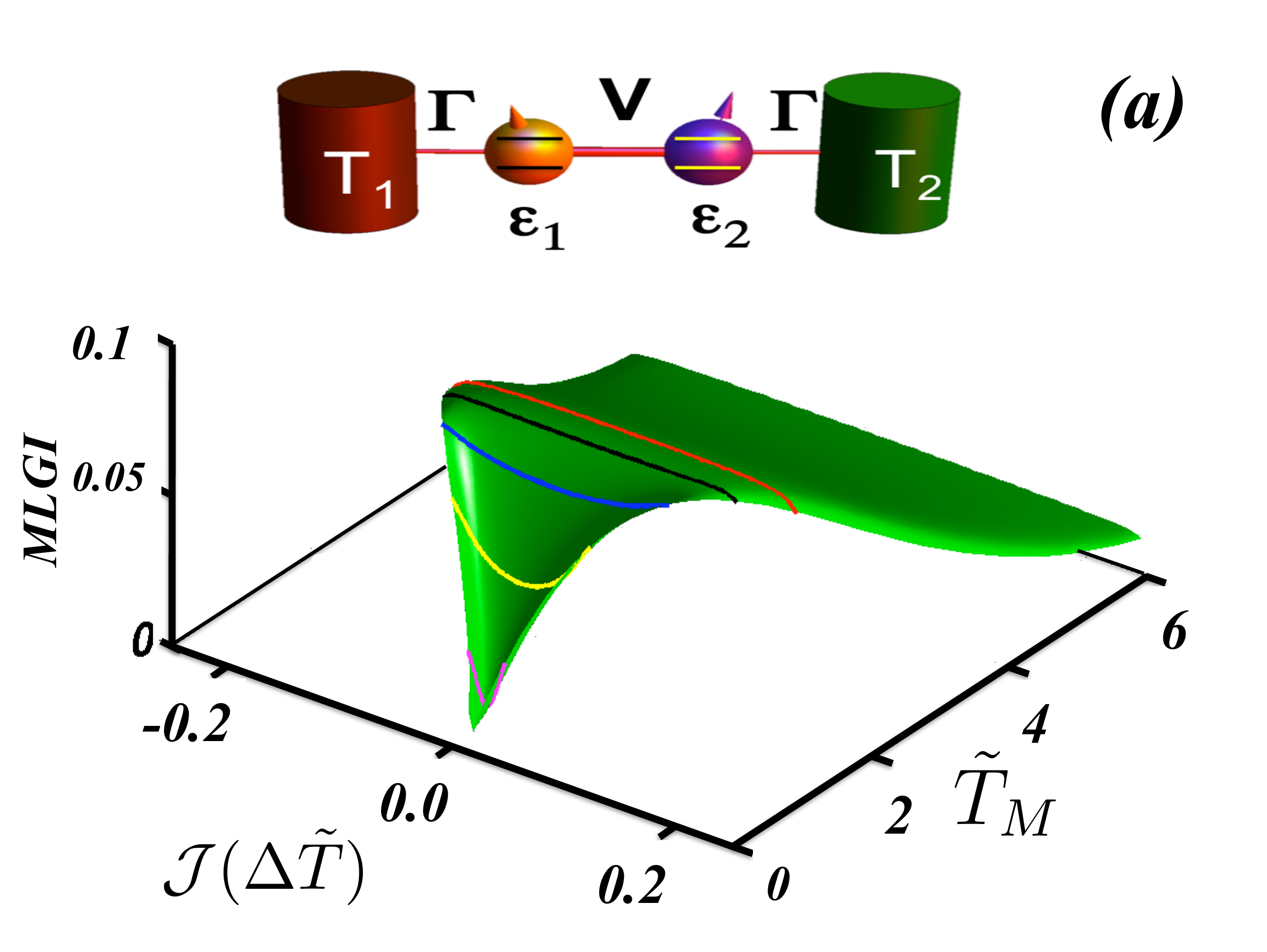}
\includegraphics[height=4cm,width=3.5cm]{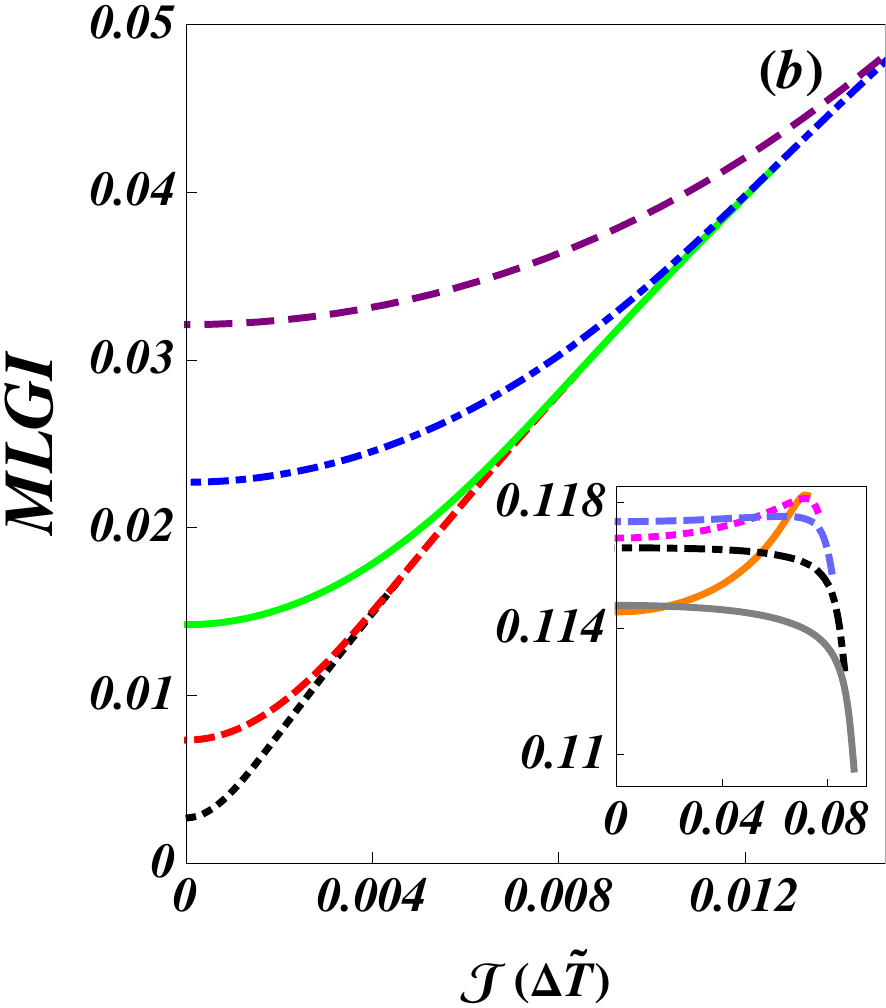}
\includegraphics[height=3.9cm,width=3.5cm]{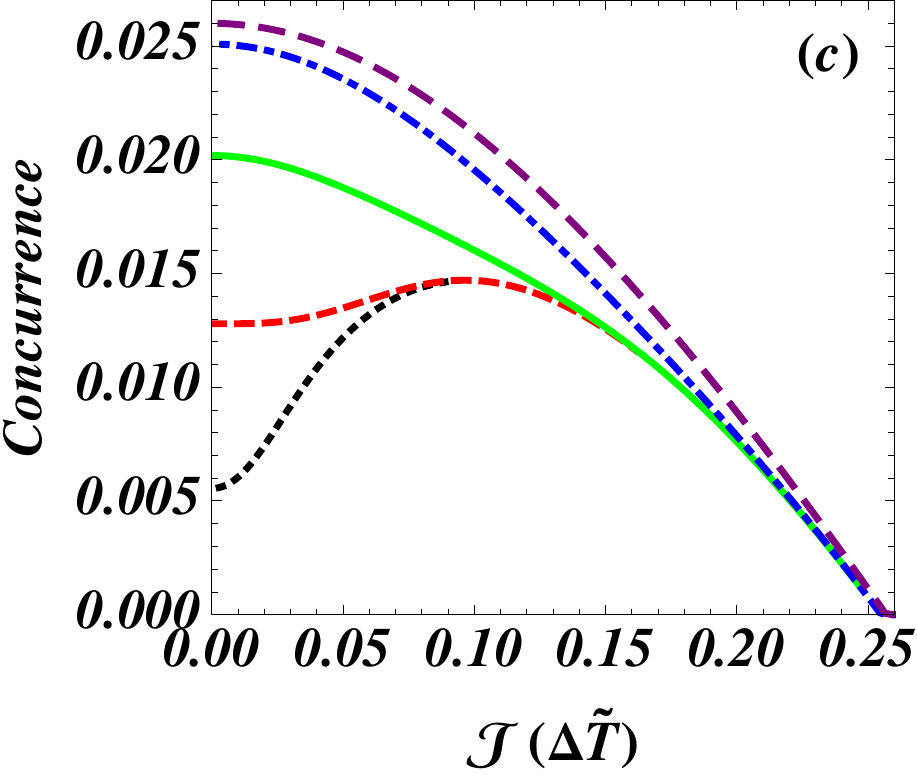}
\includegraphics[height=3.9cm,width=3.5cm]{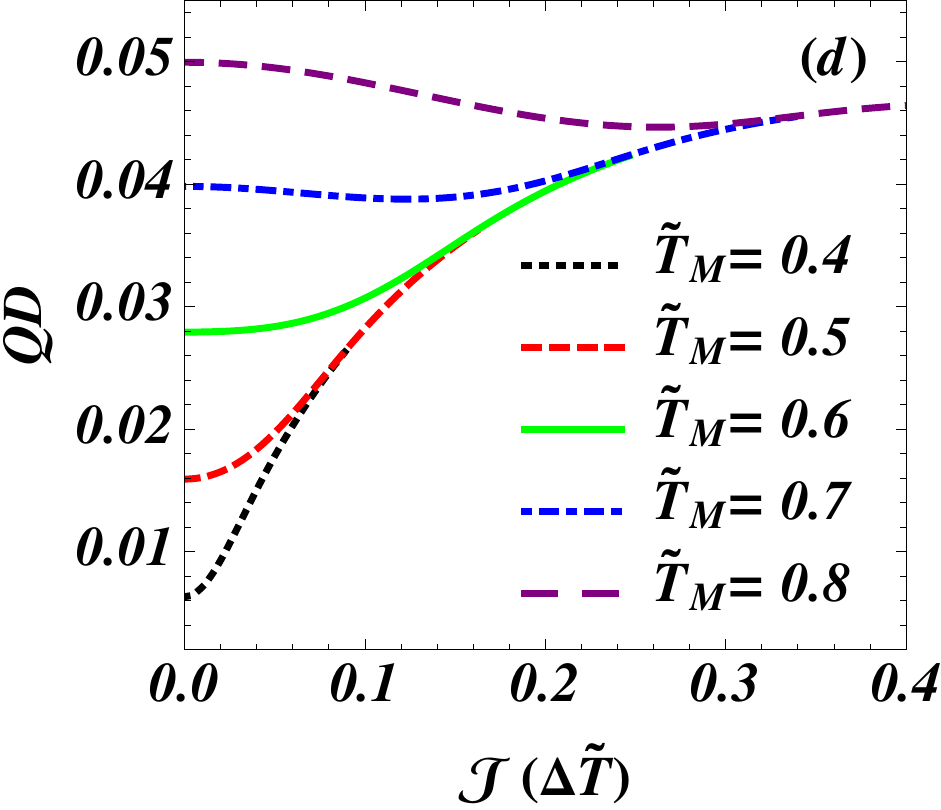}
\vspace{-0.2cm}
\caption{(Color online).
Composite quantum system in contact with two separate thermal reservoirs.
(a) A 3D plot of the {\sl first} MLGI {\sl (for $\tilde \tau=\pi/6$)} as a function of mean temperature and
heat current for a two-spin chain with $\tilde \epsilon=3$ and $\tilde \Gamma=0.05$ is shown.
(b) Slices taken from the 3D left panel are shown for different $\tilde T_M$. Notice that for the lowest values of $\tilde T_M$, MLGI increases with the application of a temperature gradient. Similar results are shown in the inset for
higher mean temperatures: Orange ($\tilde T_M=2.25$), magenta ($\tilde T_M=2.5$), purple ($\tilde T_M=2.75$),
black ($\tilde T_M=3.0$), gray ($\tilde T_M=3.25$). A qualitatively similar behavior can be observed for the two-spin chain concurrence (c) and quantum discord (d) as a function of the heat current. The color convention for curves in (b-main figure), (c) and (d) is the same. Since the values of $\tilde \epsilon$ are the same, a symmetric curve is obtained for $ \Delta \tilde T < 0.$}

\label{fig2fvst}
\end{figure}

Now, the essential question is: can the LGI violation be enhanced by a temperature gradient or equivalently by a heat current? This is analyzed by plotting in Fig. \ref{fig2fvst} (a) the MLGI as a function of ${\cal J}$ and the mean temperature $\tilde T_M=(\tilde T_1+\tilde T_2)/2$.
The variation of MLGI vs. heat current (or equivalently temperature gradient) changes its concavity indicating that the enhancement of LGI violations under nonequilibrium thermal conditions is restricted to the low mean temperature sector. This behavior can be understood as a competition of the
coherent flip-flop spin processes (dominated by the $V$ term) and a temperature dependent decay rate ($\tilde \Gamma_V$), as can be deduced from Eq. \ref{Eq:fleget}.
Since $\tilde \Gamma_V << 1$, it is possible to obtain an analytical expression for the MLGI as
$MLGI \simeq r_{23}\left [ e^{-\frac{\pi \tilde \Gamma_V}{6}}+\frac{1}{2}e^{-\frac{\pi\tilde \Gamma_V}{3}}-1\right ]$, that depends on the product of
the stationary spin chain level populations $r_{23}$ and an effective decay term.
For low $\tilde T_M$, and $\Delta \tilde T<< 2\tilde T_M$, the population term in the last analytical MLGI expression
is much more important  than the effective decay indicating that the MLGI starts to grow as the $r_{23}$ increases
for $\Delta \tilde T=0$. By contrast,  for higher $\tilde T_M$ it is very easy to demonstrate that  $r_{23}$ decreases faster than the contribution of the effective decay $\tilde \Gamma_V$ indicating that the decay term dominates and the flip-flop interaction does not contribute appreciably due to the fact that the population of each eigenstate is saturated at 1/4. This is the main reason why moving away from thermal equilibrium the MLGI decreases for high $\tilde T_M$.
The non-equilibrium enhancement of spatial quantum coherences in the same spin-chain system, as measured by the concurrence  \cite{quiroga} and QD \cite{wu1}, is shown in Fig. \ref{fig2fvst} (c)-(d), respectively. This correspondence between spatial and temporal correlations extends to a nonequilibrium thermal setting previous observations of a perfect mapping between Bell inequality and LGI \cite{reznik}.

Our method as applied to the single spin case \cite{nori,huelga}  (results not shown here) demonstrates that MLGI is not enhanced, rather it is suppressed, when a nonequilibrium temperature gradient is applied. Therefore, a two-qubit system is the smallest spin chain where nonequilibrium enhancement of LGI violations can be observed.

\begin{figure}[tbh]
\begin{center}
\includegraphics[height=5cm,width=6cm]{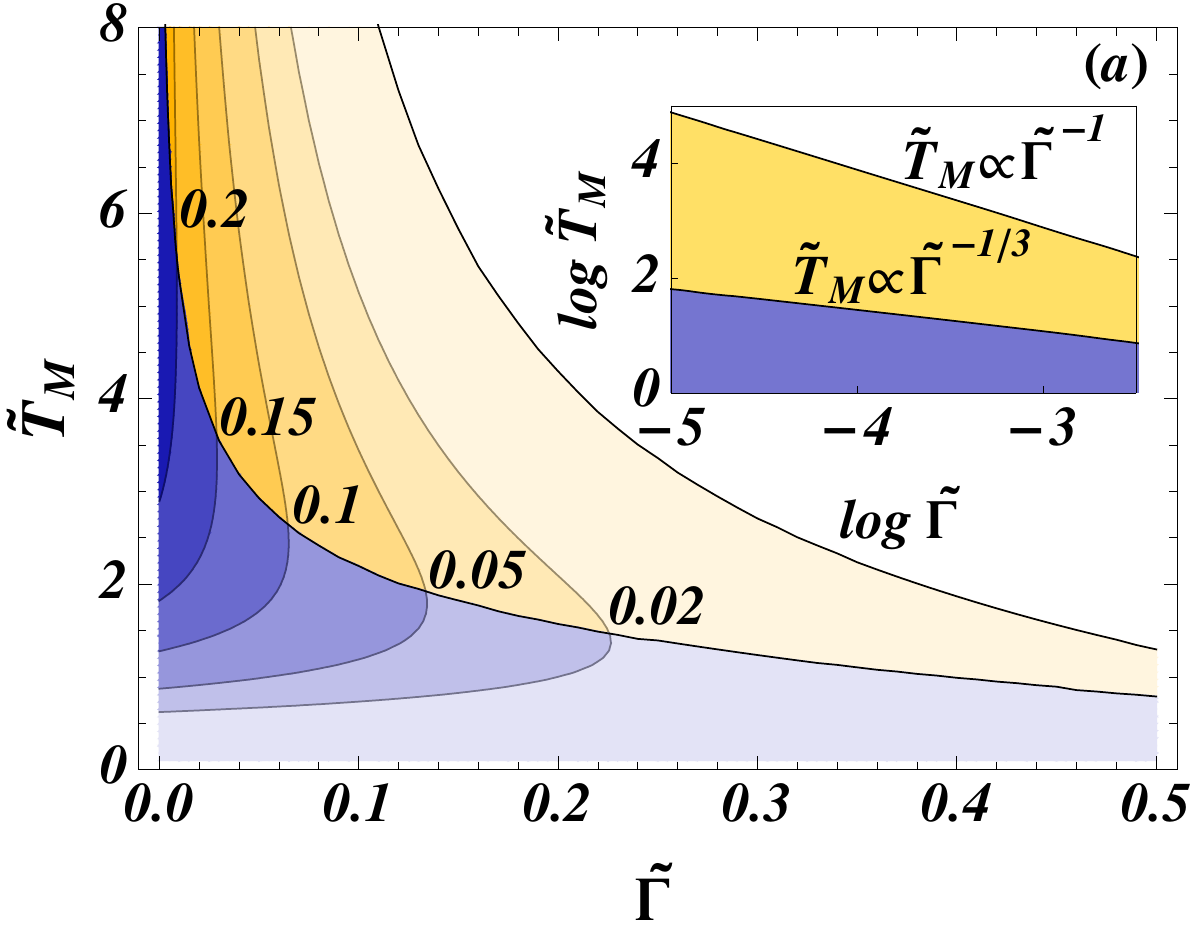}
\includegraphics[width=6cm]{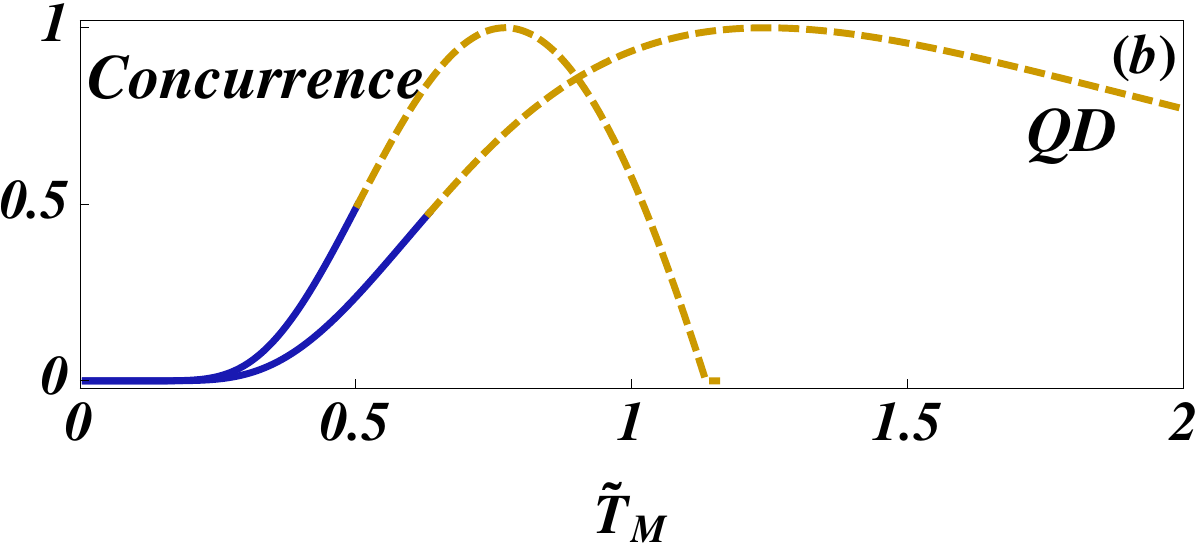}
\end{center}
 \vspace{-0.5cm}
 \caption{(Color online). (a) Contour plot of MLGI in the $(\tilde \Gamma, \tilde T_M)$ parameter space for $\tilde \epsilon=3$.
The numbers around the contour lines give the values of the MGLI at $\Delta \tilde T =0$. Inset: ($\tilde \Gamma, \tilde T_M$) in log-log
scales showing  power laws for the borderlines between different
non-equilibrium MLGI behavior regions.
(b) Typical behaviors of concurrence
and QD at equilibrium (both normalized to their maximum value), where a
blue solid line indicates that a temperature gradient increases the
quantum spatial correlation while a yellow dashed line indicates that a
temperature gradient decreases the quantum spatial correlation}
\label{diagfaseloglog}
\end{figure}

A complete presentation of LGI violations under nonequilibrium conditions is given in Fig.\ref{diagfaseloglog} (a)
We point out that thermal nonequilibrium effects on LGI violations can be grouped in three zones in the whole $(\tilde \Gamma, \tilde T_M)$ parameter space. First, there are no LGI violations, both in equilibrium and nonequilibrium, for sufficiently high $\tilde \Gamma$ or $\tilde T_M$ values, corresponding to the upper right white zone in Fig. \ref{diagfaseloglog} (a). This behavior is consistent with the deleterious effect of strong coupling or high temperature on quantum correlations. Second, there is a zone with intermediate values of $\tilde \Gamma$ and $\tilde T_M$ (yellow region in Fig.\ref{diagfaseloglog} (a) where $MLGI > 0$ at thermal equilibrium, but the MLGI starts deteriorating rather than improving
as the system is subject to a small temperature gradient. As shown in the inset, where a log-log scale has been used for $\tilde \Gamma$ and $\tilde T_M$, the numerically determined borderline between the white and yellow zones has a  power law expression such that $\tilde T_M \sim \tilde \Gamma^{-1}$. This simple relation can be easily justified by using our analytical approximation for MLGI,
by searching for the $(\tilde \Gamma, \tilde T_M)$ values yielding to a null value for MLGI at $ \Delta \tilde T = 0$. Third, and most
importantly, there is a parameter region of low $\tilde \Gamma$ or $\tilde T_M$ values (blue zone in Fig.\ref{diagfaseloglog} (a) where the equilibrium value of MLGI is further increased
as the system is taken out of equilibrium by setting a temperature
gradient. The numerically determined borderline between the yellow and blue regions in Fig.\ref{diagfaseloglog} (a) satisfies also a  power law of the form
$\tilde T_M \sim \tilde \Gamma^{-1/3}$. This behavior can also be explained by
requiring that $\frac{\partial^2 MLGI}{\partial (\Delta \tilde T)^2}|_{\Delta \tilde T=0}=0$, (details of the derivation are given in Appendix \ref{app:LGIpowlaw})
\begin{eqnarray}
\tilde \Gamma=\frac{3}{\pi}\left[\frac{\tilde \epsilon}{4}\right]^3\left[1-\left(\frac{1}{\tilde \epsilon}\right)^4\right]\frac{1}{\tilde T_M^3} \label{Eq:299}
\end{eqnarray}
in agreement with the numerical results plotted in Fig. \ref{diagfaseloglog} (a). Clearly, when the inter-qubit interaction is greater than the energy
splitting,  no solution for Eq.(\ref{Eq:299}) exists indicating that a requirement for the thermal nonequilibrium enhancement of LGI violations is that spins are in the weak coupling regime. By contrast with two-time quantum correlations, two-point spatial correlations in the steady-state do not depend on the quantum system coupling strength with the thermal baths. However, as shown in Fig.\ref{diagfaseloglog} (b), the two-spin concurrence and QD also present a nonequilibrium thermal enhancement at low enough temperatures signaling one more time the correspondence between temporal and spatial quantum correlation behaviors. As it has already been documented the QD persists up to larger temperatures as compared with the concurrence.

\section{Conclusions}
\label{sec:V}
In summary, we have provided a microscopic derivation of LGI, for a spin chain coupled to two reservoirs
at different temperatures. Nonequilibrium thermal quantum correlations have been analyzed exactly by mapping the original two-{\it interacting} spins coupled chain to an equivalent system of two-noninteracting spins, each one of them in contact with a renormalized heat reservoir at an effective temperature.
Based on a Kraus-operator approach, we demonstrate
that in a certain range of temperature gradients the steady-state LGI violation can be enhanced. The frontiers between
different behaviors of the MLGI response to thermal nonequilibrium conditions have been found to be characterized by  power laws relating mean temperature of the heat reservoirs and the
spin chain coupling strength with the baths. This leads to the interesting feature that nonequilibrium
thermal conditions provide the opportunity to enhance not only spatial but also temporal quantum correlations.
\acknowledgments
We acknowledge Susana Huelga and Neil F. Johnson's critical readings of the manuscript and Facultad de Ciencias, Vicerrector\'{\i}a de investigaciones
at Universidad de los Andes, for financial support to the project  {\it Quantum control of Non-Equilibrium hybrid systems} (2012-2014).
\appendix

\section{Time-dependent quantum correlations}
\label{app:Atimedepquancorr-details}

We start writing the expression  for $C(t_1,t_2)$, also by simplifying it only in terms of $\Pi^+$ (with $\Pi^-=I-\Pi^+$). We must do a similar expansion of all the 15 terms as the one we did for the case of unitary evolution of a pure state, after which we obtain the following expression:

\begin{align}
 C(t_1,t_2)=1-2\mathrm{Tr}\left\{\Pi^+\sum_{\nu\mu} K_\nu (t_i-t_l) K_\mu(t_l)\rho_0 K^\dagger_\mu(t_l) K^\dagger_\nu (t_i-t_l)\right\} \nonumber \\-\mathrm{Tr}\left\{\sum_{\nu\mu} K_\nu (t_i-t_l)\Pi^+ K_\mu(t_l)\rho_0 K^\dagger_\mu(t_l) K^\dagger_\nu (t_i-t_l)\right\} \nonumber \\-\mathrm{Tr}\left\{\sum_{\nu\mu} K_\nu (t_i-t_l) K_\mu(t_l)\rho_0 K^\dagger_\mu(t_l) \Pi^+ K^\dagger_\nu (t_i-t_l)\right\} \nonumber \\+2\mathrm{Tr}\left\{\Pi^+\sum_{\nu\mu} K_\nu (t_i-t_l)\Pi^+ K_\mu(t_l)\rho_0 K^\dagger_\mu(t_l) K^\dagger_\nu (t_i-t_l)\right\} \nonumber \\+2\mathrm{Tr}\left\{\Pi^+\sum_{\nu\mu} K_\nu (t_i-t_l) K_\mu(t_l)\rho_0 K^\dagger_\mu(t_l) \Pi^+ K^\dagger_\nu (t_i-t_l)\right\}
\end{align}
Note that the second term is the probability of measuring $+1$ after evolving $\rho_0$ until $t_l$ and then until $t_i$, so that it is simply $-2p(^+t_i)$. The third term can be rewritten as follows:

\begin{equation}
 \mathrm{Tr}\left\{\sum_{\nu} K_\nu (t_i-t_l)\left[\sum_{\mu}\Pi^+ K_\mu(t_l)\rho_0 K^\dagger_\mu(t_l)\right] K^\dagger_\nu (t_i-t_l)\right\}
\end{equation}
The evolution represented by the Kraus operators is trace-keeping, so that this is the same as

\begin{equation}
 \mathrm{Tr}\left\{\sum_{\mu}\Pi^+ K_\mu(t_l)\rho_0 K^\dagger_\mu(t_l)\right\}=p(^+t_l)
\end{equation}
The same happens with the fourth term, so that these two terms add up to $-2p(^+t_l)$. The last two terms can be rewritten as $2g(t_l,t_i)+h.c.$ if we define
\begin{equation}
g(t_l,t_i)=\mathrm{Tr}\left\{\rho_0 \sum_{\nu\mu} K^\dagger_\mu(t_l) K^\dagger_\nu (t_i-t_l) \Pi^+ K_\nu (t_i-t_l) \Pi^+ K_\mu(t_l) \right\}
\end{equation}
or, alternatively
\begin{align}
g(t_l,t_i)=\mathrm{Tr}\left\{\sum_{\nu\mu} \Pi^+ K_\nu (t_i-t_l) \Pi^+ K_\mu(t_l)\rho_0 K^\dagger_\mu(t_l)  K^\dagger_\nu (t_i-t_l)  \right\} \nonumber \\
=\mathrm{Tr}\left\{\Pi^+\sum_{\nu} K_\nu (t_i-t_l) \Pi^+ \rho(t_l) K^\dagger_\nu (t_i-t_l)  \right\} \label{eq:gMixto}
\end{align}

Thus, the correlations can then be written as

\begin{equation}\label{eq:correlacionMixto}
 C(t_1,t_2)=1-2p(^+t_1)-2p(^+t_2)+4\mathrm{Re}\left[g(t_1,t_2)\right]
\end{equation}
This expression depends mainly of the correlation function function $g$. It can also be shown that, replacing the Kraus operators with unitary evolution operators, we get exactly the result obtained for pure states. Based on the correlations $C$, the LGI expresion can be written as
\begin{eqnarray}
F(t_1,t_2,t_3)=C(t_1,t_2)+C(t_2,t_3)-C(t_1,t_3)\leq 1\\
F(t_1,t_2,t_3)=1-4p(^+t_2)+4Re\left[g(t_1,t_2)+g(t_2,t_3)-g(t_1,t_3)\right]\leq 1
\label{eq:apALG1General}
\end{eqnarray}

\section{Power law behaviour of border lines between different $\Gamma-T_M$ regions}
\label{app:LGIpowlaw}
First is important to remark that the violation dependes on a population transfer between the two levels: symmetric and antisymetric
$\rho_{34}$ and an effective decay $\Gamma_v$. The violation depends on the product of those terms
$V=GH$, where $G=\rho_{3,4}$ and $H$ depends on the effective decay. Before to define all important ingredients it is necessary to
gives some algebraic details for temperature:
\begin{eqnarray*}
T_M=\frac{T_1+T_2}{2} ; \Delta T=T_2-T_1\\
T_M+\frac{\Delta T}{2}=T_1 ;
T_M-\frac{\Delta T}{2}=T_2\\
T_1T_2=T_M^2-\left(\frac{\Delta T}{2}\right)^2
\end{eqnarray*}
\begin{eqnarray}
V_{max}=GH\\
G=\frac{1}{2}\left(1-\Delta_1 \Delta_2\alpha\right)\\
H=e^{-\frac{\Gamma_v\pi}{6V}}+\frac{1}{2}e^{-\frac{\Gamma_v\pi}{3V}}-1\\
\Gamma_v=2\Gamma\left({\frac{\sinh(E_1 a)}{\Delta_1}+\frac{\sinh(E_2 a)}{\Delta_2}}\right)\\
a=\frac{T_M}{T_1T_2}=\frac{T_M}{T_M^2-\left(\frac{\Delta T}{2}\right)^2}=\frac{1}{T_M}\frac{1}{1-\left(\frac{\Delta T}{2T_M}\right)^2}\\
\Delta_i=2\left(\cosh(E_i a)-\cosh\left (E_i a\frac{\Delta T}{2T_M}\right)\right)\\
\alpha=\frac{1}{4 \sinh(E_1 a)\sinh(E_2 a)}
\end{eqnarray}
where $V$ denotes the interaction qubit parameter. $E_{2,1}=\epsilon\pm V$. We are interested into calculate the curvature of
$V_{max}$, because it could be give signatures fot the power law behaviour. It is important to refer to
one of the main results of LGI, as is Figure 4. The power law behaviour $\Gamma$ vs $T$ comes from the fact that a inverse power law
 that can be obtained when the violation is zero. This calculation is very simple and we do not show here. The main finding of
 our work is the power law $T_M^{-3}$. On the
same footing comes from a equilibrium case, where the violations start to decrease , thus, the violations
changes its curvature. Therefore it is necessary in both cases to calculate the $V_{max}$ and $V_{max}''$ behaviours in the
limit of $\Delta T=0$. As a first step we will
calculate the
 second derivative with respect to $\Delta T$ and we will take the limit where
$\Delta T\rightarrow 0$.
\begin{eqnarray}
\frac{d^2 V_{max}}{d\Delta T^2}=
G''H+GH''\\
\end{eqnarray}
Where 
the first derivatives gives a zero value in the maximum.
First, we need to calculate the second derivative of $H$:
\begin{eqnarray}
\frac{d^2H}{d\Delta T^2}&=&
-\frac{\pi}{6V}\left[e^{-\frac{\Gamma_v\pi}{6V}}+e^{-\frac{\Gamma_v\pi}{3V}}\right]\frac{d^2\Gamma_v}{d\Delta T^2}
\end{eqnarray}
Next we have to calculate the $\Gamma_V^{''}$ term.

\begin{eqnarray}
\frac{d^2\Gamma_v}{d\Delta T^2}&=&
2\Gamma a^2
\left\{
\frac{E_1}{2T_M}
\left(
a\frac{\Delta T }{T_M}\frac{\Delta T}{\Delta_1}-\frac{\Delta T}{\Delta_1^2}\frac{d\Delta_1}{d\Delta T}+\frac{1}{\Delta_1}
\right)\cosh(E_1 a)
+\frac{(E_1 a \Delta T)^2}{4\Delta_1 T_M^2}\sinh(E_1 a)-
\right .
\\
& &
\left .
 \frac{E_1}{2}\frac{\Delta T}{T_M}\frac{d\Delta_1}{d\Delta T}\frac{1}{\Delta_1^2}\cosh(E_1 a)+
\frac{2\sinh(E_1a)}{\Delta_1}\left(\frac{1}{a\Delta_1}\frac{d\Delta_1}{d\Delta T}\right)^2-\frac{\sinh(E_1 a)}{(a\Delta_1)^2}\frac{d^2\Delta_1}{d\Delta T^2}
\right .
\\
& &
\left .
+ 1\rightarrow 2\frac{}{}
\right\}\\
\end{eqnarray}
Second, we calculate de second derivative of G
\begin{eqnarray}
G''=-\frac{1}{2}(\Delta_1''\Delta_2\alpha+\Delta_1\Delta_2''\alpha+\Delta_1\Delta_2\alpha'')
\end{eqnarray}
Next we have to calculate de second derivative of $\alpha$
\begin{eqnarray}
\frac{d^2\alpha}{d\Delta T^2}&=&
-\frac{a^2}{T_M^2}
\left\{
\frac{(\alpha a  \Delta T)^2}{2 T_M}
\left [
(E_1^2 + E_2^2)\sinh(E_1 a) \sinh(E_2 a)+2E_1E_2 \cosh(E_2 a) \cosh(E_1 a)
\right]
\right .\\
&-& \frac{T_M}{a} \frac{d\alpha}{d\Delta T}\left (\frac{2 T_M}{\alpha a}\frac{d\alpha}{d\Delta T}+\Delta T\right)
\left .
+2 \alpha^2 T_M
\left[E_1 \cosh(E_1 a) \sinh(E_2 a)+E_2 \cosh(E_2 a) \sinh(E_1 a)\right]
\right\}
\end{eqnarray}

The point of interest is to take the limit when $\Delta T=0$ for the second derivatives of $\Delta_i$ and $\Gamma$.
\begin{eqnarray}
\frac{d^2\Delta_i}{d\Delta T^2}|_{\Delta T=0}
&=&
a^2
\left[\frac{E_i }{T_M}\sinh(E_i a)
-\frac{(E_i a)^2}{2}\frac{1}{aT_M}\cosh\left(\frac{E_i a \Delta T}{2T_M}\right)
\right]\\
&=&
\frac{a^2 E_i}{T_M}
\left[\sinh(E_i a)
-\frac{E_i a}{2}
\right]=
\frac{E_i}{T_M^3}
\left[\sinh\left(\frac{E_i}{T_M}\right)
-\frac{E_i}{2 T_M}
\right]\\
\frac{d^2\Gamma_v}{d\Delta T^2}|_{\Delta T=0}&=&
2\Gamma a^2
\left\{
\frac{E_1}{2T_M\Delta_1}\cosh(E_1 a)-\frac{\sinh(E_1 a)}{(a\Delta_1)^2}\frac{d^2\Delta_1}{d\Delta T^2}
+ 1\rightarrow 2\frac{}{}
\right\}\\
&=&
\frac{\Gamma} {2T_M^2}
\left\{
\frac{E_1}{T_M}\frac{\cosh(\frac{E_1} {T_M})}{\cosh(\frac{E_1} {T_M})-1}+
\frac{E_1}{T_M}\frac{\sinh(\frac{E_1} {T_M})}{(\cosh(\frac{E_1} {T_M})-1)^2}\left[\frac{E_1}{2T_M}-\sinh\left(\frac{E_1} {T_M}\right)\right]
+ 1\rightarrow 2\frac{}{}
\right\}\\
\end{eqnarray}
With those ingredientes calculated before and 
After a long algebra and defining $x_i=\frac{E_i}{2T_M}$, $G''$, in the limit where $\Delta T=0$, can be written as:
\begin{eqnarray}
H''=-\frac{\pi}{3V}\left[1-\frac{\Gamma\pi}{2V}\right]\frac{\Gamma}{2T_M^2}\sum_{i=1}^2x_i \left[x_i-\tanh(x_i)\right ]\frac{\cosh(x_i)}{\sinh^3(x_i)}
\end{eqnarray}
To calculate the second derivative is necesary to evaluate $G$ and $H$ at $\Delta T=0$. First $G$:
\begin{eqnarray}
G|_{\Delta T=0}&=&\frac{1}{2}\left(1-\Delta_1 \Delta_2\alpha\right)\\
&=&\frac{1}{2}\left[1-\tanh(x_1)\tanh(x_2)\right]\\
\end{eqnarray}
Second $H$. To calculate it, we need $\Gamma_v|_{\Delta T=0}$. We assume the limit of low average temperatures
\begin{eqnarray}
\Gamma_v|_{\Delta T=0}
&\approx& 2\Gamma\\
H|_{\Delta T=0}&=&e^{-\frac{\Gamma_v\pi}{6V}}+\frac{1}{2}e^{-\frac{\Gamma_v\pi}{3V}}-1\\
&=&\frac{1}{2}\left(1-\frac{4\Gamma\pi}{3V}\right)
\end{eqnarray}
The main objective is to calculate the second derivative of $V_{max}$, when $\Delta T\rightarrow 0$. We define $x_i=\frac{E_i}{T_M}$ and
To explain the concavity we need:
\begin{eqnarray}
\frac{d^2V}{d\Delta T^2}&=&G''H+GH''\\
\end{eqnarray}
Clearly the second derivative of $H$ is always negative, allowing that we can find $V_{max}$ can be zero.
We show a plot of both curvatures
\begin{figure}[tbh]
\includegraphics[height=5.5cm,width=8.8cm]{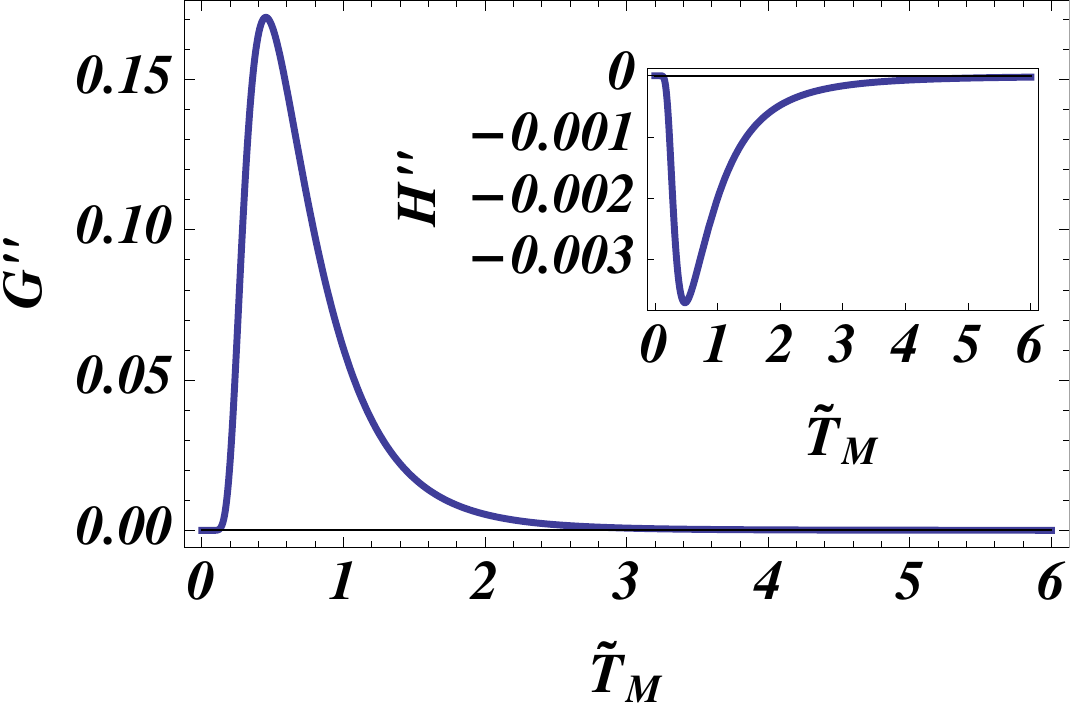}
\includegraphics[height=5.6cm,width=7.8cm]{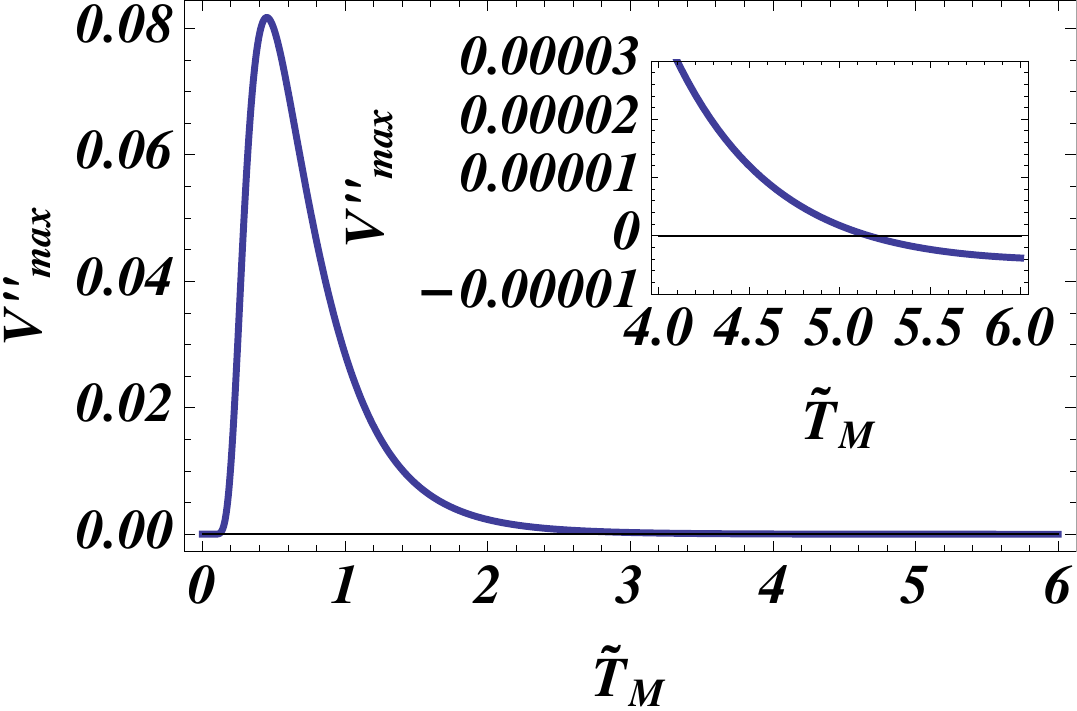}
\caption{Second derivatives of $V_{max}$, G and H with respect to $\Delta T$, for $\tilde \epsilon=3$ and $\tilde \Gamma=0.01$}
\label{fig:curvaturas}
\end{figure}
This means that the second derivative of the maximum violation depends on which effect is stronger between the convexity of $G$ and the concavity of $H$. We therefore show plots of both curvatures and their sum in figure \ref{fig:curvaturas}, along with a plot of the second derivative of $V_{max}$.  Fig.\ref{fig:curvaturas} shows that the second derivative of $G$ is always positive, whereas the second derivative of $H$ is always negative. The result is that the second derivative of $V_{max}$ depends on which of these two effects is prevalent. For low $T_m$, $G''$ increases very fast, so that $V''_{max}$ is positive. However, $G''$ also decays faster than $H''$ with high $T_m$, so that the fact that $H''<0$ is prevalent for high mean temperatures. We can see this in Fig.\ref{fig:curvaturas},
left panel, where the inset shows that the second derivative crosses the horizontal axis at around $T_m=5$.

The conclusion is that for low temperatures the effect of $G$ is much more important than the effect of H. This is the main reason for the increased violations out of equilibrium.  On the other hand, as the temperature increases the effect of $G$ vanishes, which is not surprising since the population of each eigenstate tends to $\frac{1}{4}$, while the effect of $H$ remains. This is the reason why moving away from equilibrium reduces the maximum violation to the LGI for high temperatures.

To justify the numerical calculations, we approximate at first order $\tanh(x)=\frac{e^x-e^{-x}}{e^x+e^{-x}}\approx\frac{2x}{2+x^2}$, that is valid for the limit of $|\frac{E_i}{2T_M}|<1$.
Additionally, we take into account the $\sinh(x)\propto x$ and $\cosh(x)\propto 1+x^2/2$
As we are working in the Markovian limit and to be consistent with the linblad approximation, terms proportional to $\Gamma^2$, goes to zero.
After a long algebra we get the following equation
\begin{eqnarray*}
x_1^3 x_2+x_2^3 x_1&=&\frac{2\pi}{3}\frac{\Gamma}{V}(x_1+x_2)(4+2(x_1^2+x_2^2)+x_1x_2(x_1x_2-1))\\
&\approx&\frac{8\pi}{3}\frac{\Gamma}{V}(x_1+x_2)\\
\frac{E_1^3 E_2}{T_M^4}+\frac{E_2^3 E_1}{T_M^4}&=&\frac{4^3\pi}{3}\frac{\Gamma}{V}(\frac{E_1}{T_M}+\frac{E_2}{T_M})\\
\frac{3}{4^3\pi}\frac{E_1E_2}{E_1+E_2}(E_1^2+E_2^2)\frac{1}{T_M^3}&=&\frac{\Gamma}{V}
\end{eqnarray*}
By normalizing to the interaction term $V$
\begin{eqnarray}
\Gamma=\frac{3}{4^3\pi}\frac{\epsilon^4-1}{\epsilon}\frac{1}{T_M^3}
\end{eqnarray}

\end{document}